\documentclass[12pt]{article} 
\pdfoutput=1
\usepackage{amssymb,graphicx}
\usepackage{epstopdf}
\usepackage{amsmath,amsfonts}
\usepackage{epsfig} 
\usepackage{graphicx,graphics}
\usepackage{physics}
\usepackage{color}
\usepackage{soul}
\usepackage[unicode=true,pdfusetitle,
 bookmarks=true,bookmarksnumbered=true,bookmarksopen=true,bookmarksopenlevel=3,
 breaklinks=false,pdfborder={0 0 1},backref=false,colorlinks=true,citecolor=blue]
 {hyperref}

\begin{document} 

\sloppy
\title{Shimmering gravitons in the gamma-ray sky}

\author{S. Ramazanov$^{a}$, R. Samanta$^{a}$, G. Trenkler$^{a, b}$, F. Urban$^{a}$\\
\small{\em ${}^{a}$CEICO, Institute of Physics of the Czech Academy of Sciences,}\\
\small{\em Na Slovance 1999/2, 182 00 Prague 8, Czech Republic}\\
\small{\em ${}^{b}$Institute of Theoretical Physics, Faculty of Mathematics and Physics, Charles University, }\\
\small{\em V Holešovičkách 2, 180 00 Prague 8, Czech Republic}}

{\let\newpage\relax\maketitle}
\begin{abstract}

What is the highest energy at which gravitons can be observed? We address this question by studying graviton-to-photon conversion---the inverse-Gertsenshtein effect---in the magnetic field of the Milky Way. We find that above $\sim 1~\mbox{PeV}$ the effective photon mass grows large enough to quench the conversion rate. For sub-PeV energies, the induced photon flux is comparable to the sensitivity of LHAASO to a diffuse \(\gamma\)-ray background, but only for graviton abundances of order $\Omega_{\text{gw}} h^2_0 \sim 1$. In the future, owing to a better understanding of \(\gamma\)-ray backgrounds, larger effective areas and longer observation times, sub-PeV shimmering gravitons with a realistic abundance of $\Omega_{\text{gw}} h^2_0 \sim 0.01$ could be detected. We show how such a large abundance is achieved in a cosmologically-motivated scenario of post-recombination superheavy dark matter decay. Therefore, the sub-PeV range might be the ultimate energy frontier at which gravitons can be observed.

\end{abstract}

% --  --  --  --  --  --  --  --  --  --  --  --  --  --  --  --  --  --  --  --  --  --  --  --  --  -- -
\section{Introduction}\label{sec:intro}
% --  --  --  --  --  --  --  --  --  --  --  --  --  --  --  --  --  --  --  --  --  --  --  --  --  -- -

The conversion of photons to gravitons (and vice versa) in a strong static electromagnetic field---the (inverse) Gertsenshtein effect~\cite{Gertsenshtein}---enables the testing of gravity in the high-frequency regime, well beyond the reach of ground- or space-based interferometers. This phenomenon is analogous to axion-photon conversion~\cite{Sikivie:1983ip}, but it does not require new physics: it is built-in in the action of electromagnetism due to the inevitable coupling of the electromagnetic tensor $F_{\mu \nu}$ to the metric $g_{\mu \nu}$. The graviton-to-photon conversion is strongly suppressed because it is a gravitational process, but the suppression can be partially compensated in astrophysical and cosmological backgrounds thanks to large distances traversed by a photon or a graviton, which is the setting we consider in this work\footnote{See Ref.~\cite{Aggarwal:2020olq} for a review of the broad range of facilities and environments in which high-frequency gravitons can be searched for.}. So far, the study of high-frequency gravitons in cosmology has been mainly limited to the MHz -- GHz frequency range characteristic of $21~\mbox{cm}$ and CMB physics~\cite{Chen:1994ch, Pshirkov:2009sf, Domcke:2020yzq}. The impact of gravitons with frequencies up to $10^{18}~\mbox{Hz}$ (corresponding to energies $\sim 10~\mbox{keV}$) on the cosmic X-ray background has been investigated in Refs.~\cite{Dolgov:2012be, Dolgov:2013pwa}. In this work, we pursue this idea to its extreme and address the question of what is the highest energy at which a cosmological abundance of gravitons can be detected. We show that this energy is in the sub-PeV range covered by LHAASO~\cite{Ma:2022aau} and marginally by the Cherenkov Telescope Array (CTA)~\cite{CTAConsortium:2010umy,CTAConsortium:2017dvg}.

This result stems from the combination of two observations. Firstly the propagation length of \(\gamma\)-rays is rather short, shrinking to $1~\mbox{Mpc}\text{--}10~\mbox{kpc}$ for energies of $100~\mbox{TeV}\text{--}\mbox{1~\mbox{PeV}}$~\cite{Dwek:2012nb}. For this reason we can limit our analysis to the Milky Way only, as graviton-to-photon conversion happening further away would be unobservable from Earth. Secondly, in a magnetic field, the photon acquires an effective mass that grows linearly with frequency. This effect reduces the graviton-to-photon conversion probability above a certain cutoff. In the case of the Milky Way's magnetic field this cutoff is approximately $\omega \sim 1~\mbox{PeV}$, see Sec.~\ref{sec:gert}.

We model the magnetic field of the Milky Way according to Refs.~\cite{GMF, Pshirkov:2011um}. While these models disagree on the details of the magnetic field's spatial shape, the predicted maximal \(\gamma\)-ray flux is consistent within a factor of~2. We find that the \(\gamma\)-ray flux is comparable to the sensitivity of LHAASO only for a relic abundance of gravitons of $\Omega_\text{gw} h^2_0 \sim 1$ (Sec.~\ref{sec:flux}). A graviton abundance this high is excluded, but future improvements in the experimental sensitivity and better modelling of the \(\gamma\)-ray background will allow to test realistic scenarios. In Sec.~\ref{sec:dm} we describe one such scenario, in which gravitons with a relic abundance as large as $\Omega_\text{gw} h^2_0 \sim 0.01$ are produced sufficiently late, i.e., after recombination, through the decay of superheavy dark matter~\cite{Ema:2021fdz} motivated in the context of Hubble~\cite{Berezhiani:2015yta} and $\sigma_8$~\cite{Enqvist:2015ara} tensions.

% --  --  --  --  --  --  --  --  --  --  --  --  --  --  --  --  --  --  --  --  --  --  --  --  --  -- -
\section{The inverse Gertsenshtein effect}\label{sec:gert}
% --  --  --  --  --  --  --  --  --  --  --  --  --  --  --  --  --  --  --  --  --  --  --  --  --  -- -

We review here the Gertsenshtein effect and emphasise the aspects relevant for a very-high-energy regime and for the propagation in an inhomogeneous magnetic field---the analogous case of axion-photon oscillations can be found in Refs.~\cite{Raffelt:1987im, Mirizzi:2007hr, Simet:2007sa, Mirizzi:2009aj, Dobrynina:2014qba, Kartavtsev:2016doq}. Throughout the paper, we use natural Lorentz-Heaviside units $\hbar=c=1$ and the Minkowski metric with signature $\eta_{\mu\nu}=\text{diag}[1, -1, -1, -1]$.

The combined evolution of gravitons and photons is described by the action:
\begin{equation}
\label{GREM}
S_{\text{GR+EM}}=- \int d^4 x\sqrt{-g} \left[ \frac{M^2_\text{P}}{2} R+\frac{1}{4} g^{\mu \lambda} g^{\nu \rho} F_{\mu \nu} F_{\lambda \rho}+e\bar{\Psi} \gamma^{\mu} \Psi A_{\mu} \right]  \; ,
\end{equation}
where $M_\text{P} \approx 2.45 \cdot 10^{18}~\mbox{GeV}$ is the reduced Planck mass, $F_{\mu \nu}=\partial_{\mu} A_{\nu}-\partial_{\nu} A_{\mu}$ is the electromagnetic strength tensor, $\Psi$ are spinors describing electrons and positrons, and $e$ is the electromagnetic coupling constant. For most of the processes discussed below, we can integrate out the fermions and replace the fermion interaction with the effective Euler-Heisenberg Lagrangian: 
\begin{equation}
\label{EH}
e\bar{\Psi}\gamma^{\mu} \Psi A_{\mu} \rightarrow \frac{\alpha^2}{90 m^4_\text{e}} \left[\left(F_{\mu \nu} F^{\mu \nu} \right)^2 +\frac{7}{4} \left(\tilde{F}_{\mu \nu} F^{\mu \nu} \right)^2 \right] \; ,
\end{equation}
where $m_\text{e} \approx 0.511~\mbox{MeV}$ and $\alpha \approx 1/137$ are the electron mass and fine-structure constant, respectively. The validity of the Euler-Heisenberg Lagrangian is limited to processes with centre-of-mass energies lower than the electron mass, which is the case throughout this work unless otherwise specified.

Graviton-photon mixing in a strong magnetic field is encoded in the coupling of the metric tensor $g_{\mu \nu}$ to the electromagnetic strength tensor $F_{\mu \nu}$. We consider a randomly oriented magnetic field. In the Coulomb gauge, a photon with energy $\omega$ is described by 
\begin{equation}\label{Aexp}
A_{i} (\vec{x}, t) =\sum_{\lambda=\parallel, \bot} A_{\lambda} (\vec{x})  {\epsilon}_{i}^{\lambda} e^{-i\omega t} \; ,
\end{equation}
while a graviton is described by the spin-2 field 
\begin{equation}\label{hexp}
h_{ij} (\vec{x}, t)=\sum_{\lambda =\times, +} h_{\lambda} (\vec{x}) e^{\lambda}_{ij} e^{-i\omega t} \; .
\end{equation}
In Eq.~\eqref{Aexp}, $\epsilon_{i}^{\lambda}$ with $\lambda =\{ \parallel, \bot \}$ are the photon polarisation vectors, written as a pair of orthonormal vectors that are orthogonal to the direction of propagation, which we choose to be the $z$-axis.
The spin-2 polarisation tensors $e^{\lambda}_{ij}$ entering in Eq.~\eqref{hexp} are chosen as follows:
\begin{equation}
e^{\times}_{ij}=\epsilon^{\parallel}_i \epsilon^{\bot}_j+\epsilon^{\bot}_i \epsilon^{\parallel}_j \qquad e^{+}_{ij}=\epsilon^{\parallel}_i \epsilon^{\parallel}_j-\epsilon^{\bot}_i \epsilon^{\bot}_j \; .
\end{equation}
In the regime for which the frequency $\omega$ changes very slowly on time scales $\sim \omega^{-1}$, which applies in our case, the equation describing the change of the mode functions $A_{\lambda}$ and $h_{\lambda}$ can be written as
\begin{equation}
\label{eom}
\left(i\frac{d}{dz} +\omega \right)\begin{pmatrix} h_{+}\\ h_{\times}\\ A_{\parallel} \\ A_{\bot}\end{pmatrix}=\mathcal{H}\begin{pmatrix} h_{+}\\ h_{\times}\\ A_{\parallel} \\ A_{\bot} \end{pmatrix} \; .
\end{equation}
The mixing Hamiltonian $\mathcal{H}$ has the structure: 
\begin{equation}
\label{H0+andx}
\mathcal{H} =\begin{pmatrix} 0 & C_{{\text g} \gamma} \\  C^{\dagger}_{{\text g}\gamma}  & C_{\gamma \gamma} \end{pmatrix} \; .
\end{equation}
Here $C_{\gamma \gamma}$ and $C_{{\text g} \gamma}$ are $2\times 2$ matrices. The former is a mass matrix of a photon propagating through a medium, while $C_{{\text g}\gamma}$ describes the graviton-photon mixing. The latter is the most relevant one for our discussion. It is given by 
\begin{equation}
\label{ggamma}
C_{{\text g} \gamma} =\frac{i}{\sqrt{2}M_\text{P}} \begin{pmatrix} \vec{B} \cdot \vec{\epsilon}_{\bot} & \vec{B} \cdot \vec{\epsilon}_{\parallel} \\  -\vec{B} \cdot \vec{\epsilon}_{\parallel}  & \vec{B} \cdot \vec{\epsilon}_{\bot}\end{pmatrix}\; .
\end{equation}
See Ref.~\cite{Ejlli:2018hke} and Appendix~\ref{app:details_conv} for the details. Notice that by assuming ${\cal H}$ to be a hermitian matrix we neglect absorption of photons by extragalactic background light.  This is justified because we limit ourselves to Galactic scales, and the photon propagation length exceeds $\sim 10~\mbox{kpc}$ at all energies.

%This change might sound better: (previous) This is plausible, as soon as we limit ourselves to galactic scales, because the photon propagation length exceeds $\sim 10~\mbox{kpc}$ at all the energies $\rightarrow$ (new)

Our goal is to find the conversion probability of a graviton into a photon after traversing a distance $L$:
\begin{equation}
\label{conversiongen}
P_{{\text g} \rightarrow \gamma} =\sum_{\lambda=\parallel, \bot} \left|\langle A_{\lambda} (L)| h_{\times, +}(0)\rangle \right|^2 \; ,
\end{equation}
where the summation is performed over both photon polarisations. First, let us assume that the effective photon masses have a negligible effect on the conversion probability and set the matrix $C_{\gamma \gamma}$ to zero: $C_{\gamma \gamma}={\bf 0}_{2\times 2}$. Assuming also $B L \ll M_\text{P}$, which holds in galaxies, we obtain the conversion probability for a graviton~\cite{Mirizzi:2007hr}:
\begin{equation}
\label{simplified}
P_{{\text g} \rightarrow \gamma} (L) \approx \frac{\left|\int^{L}_0 dz'~\vec{B}_\text{T} \right|^2}{2M^2_\text{P}} \; ,
\end{equation}
where $\vec{B}_\text{T}$ is the transverse component of the magnetic field, 
\begin{equation}
\vec{B}_\text{T} \equiv \left(\vec{B} \cdot  \vec{\epsilon}_{\parallel}\right) \vec{ \epsilon}_{\parallel}+\left(\vec{B} \cdot  \vec{\epsilon}_{\bot}\right) \vec{ \epsilon}_{\bot} \; .
\end{equation}
As we will see shortly, this expression holds in a broad range of energies and magnetic field strengths. In particular, it is suitable to describe the conversion into \(\gamma\)-rays in the Milky Way.

The non-zero effective photon mass, $C_{\gamma \gamma}$ plays a crucial role above a certain energy, leading to the reduction of the graviton-to-photon conversion probability. The matrix $C_{\gamma \gamma}$ can be written as~\cite{Mirizzi:2009aj} 
\begin{equation}
\label{gammagamma}
C_{\gamma \gamma} =\begin{pmatrix} \Delta^{\parallel}_{\gamma \gamma} c^2_{\phi} +\Delta^{\bot}_{\gamma \gamma} s^2_{\phi} & \left(\Delta^{\parallel}_{\gamma \gamma}-\Delta^{\bot}_{\gamma \gamma} \right)c_{\phi} s_{\phi} \\  \left(\Delta^{\parallel}_{\gamma \gamma}-\Delta^{\bot}_{\gamma \gamma} \right)c_{\phi} s_{\phi}  & \Delta^{\parallel}_{\gamma \gamma} s^2_{\phi} +\Delta^{\bot}_{\gamma \gamma} c^2_{\phi} \end{pmatrix} \; .
\end{equation}
The $\Delta^{\parallel, \bot}_{\gamma \gamma}$ are the masses of photons with different polarisations. The angle $\theta$ is related to the direction of the magnetic field, i.e., $c_{\theta} \equiv\vec{B}\cdot\vec{ \epsilon}_{\parallel}/B_\text{T}$. Since we consider an inhomogeneous magnetic field, the off-diagonal components of the \(C_{\gamma\gamma}\) matrix effectively average out, so that it takes the form: 
\begin{equation}\label{ggapprox}
C_{\gamma \gamma} \approx \Delta_{\gamma \gamma} \hat{{\bf I}}_{2\times 2} \; ,
\end{equation}
where $\hat{{\bf I}}_{2\times 2}$ is a unit $2\times 2$ matrix; see Refs.~\cite{Ejlli:2018hke, Ejlli:2020fpt} for a more general study. The quantity $\Delta_{\gamma \gamma}$ can be split into the sum of three different contributions:
\begin{equation}
\label{DeltaOrth}  
\Delta_{\gamma \gamma}=\Delta_{\text{pl}}+\frac{11}{4}\Delta_{\text{QED}}+\Delta_{\text{CMB}} \; ,
\end{equation}
see Table~\ref{Table}. Here $\Delta_{\text{pl}}$ is due to the interaction of photons with electrons present in a given astrophysical medium. In the low-energy regime this term gives the dominant contribution~\cite{Dolgov:2017bpj}. However, in the high-energy regime we are interested in, the effects of QED vacuum birefringence $\sim \Delta_{\text{QED}}$ and the interaction with CMB photons $\Delta_\text{CMB}$ are dominant.

Note that for energies $\omega \gtrsim 100~\mbox{TeV}$, the Euler-Heisenberg approximation used to evaluate $\Delta_\text{CMB}$ ceases to be valid. Namely, photons with these energies are capable of producing electron-positron pairs by scattering off the CMB. This stops the linear growth \(\Delta_\text{CMB}\) with $\omega$, which is in fact bounded from above as
\begin{equation}
\label{CMBsmall}
\Delta_\text{CMB} \cdot 10~\mbox{kpc} \lesssim 1  \,,
\end{equation}
for all the energies, see Refs.~\cite{Dobrynina:2014qba,Mastrototaro:2022kpt}.

\begin{table}[]
\centering
\begin{tabular}{ |c|c|c| } 
 \hline
$\Delta$ & Expression \cite{Raffelt:1987im,Dobrynina:2014qba,Latorre:1994cv}  & Numerical value \\ 
\hline
$\Delta_{\text{pl}}$ & $-\frac{\omega_{\text{pl}}^{2}}{2\omega}$ & $-1.1\cdot10^{-10}\cdot\left(\frac{\omega}{1\text{TeV}}\right)^{-1}\cdot\left(\frac{n_{\text{e}}}{10^{-3}\text{cm}^{-3}}\right)\cdot\text{kpc}^{-1} $ \\ 
\hline
$\Delta_{\text{CMB}}$ & $\frac{44 \pi^2\alpha^{2}}{2025}~\frac{T^{4}_{\text{CMB}}}{m^{4}_{\text{e}}}~\omega$ & $8\cdot10^{-5}\cdot\left(\frac{\omega}{1\text{TeV}}\right)\cdot\text{kpc}^{-1} ~~~\omega \lesssim 100~\mbox{TeV}$ \\ 
\hline
$\Delta_{\text{QED}}$ & $\frac{\alpha}{45\pi}\left(\frac{B_{T}}{B_{\text{cr}}}\right)^{2}\omega$ & $1.5\cdot10^{-4}\cdot\left(\frac{\omega}{1\text{TeV}}\right)\cdot\left(\frac{B_{\text{T}}}{6 \mu\text{G}}\right)^{2}\cdot\text{kpc}^{-1}$ \\
\hline
\end{tabular}
\caption{Different contributions to the photon mass matrix. We used the expression $\omega_{\text{pl}}=n_{e}e^{2} / m_{\text{e}}$ for the plasma frequency and $B_{\text{cr}}=m^{2}_{\text{e}} / e$ for the critical magnetic field strength; $n_\text{e}$ is the electron number density, and $T_\text{CMB} \approx 2.73~\mbox{K}$ is the CMB temperature.}
\label{Table}
\centering
\end{table}

To account for the effective photon mass $\Delta_{\gamma \gamma} \neq 0$, we modify the conversion probability~\eqref{conversiongen} as follows:
\begin{equation}
\label{conversionfull1}
P_{{\text g} \rightarrow \gamma} (L) \approx \frac{\left|\int^{L}_0 dz'~ e^{i\Delta_{\gamma \gamma} z'}\vec{B}_\text{T}  \right|^2}{2M^2_\text{P}} \; .
\end{equation}
Here $\Delta_{\gamma \gamma}$ is assumed to be constant. The oscillation factor $\sim e^{i \Delta_{\gamma \gamma} z'}$ in the integrand of Eq.~\eqref{conversionfull1} is negligible, provided that 
\begin{equation}
\label{neglect}
\Delta_{\gamma \gamma } L_\text{corr} \lesssim \pi \; ,
\end{equation}
where $L_\text{corr}$ is the correlation length of the magnetic field. Since the CMB contribution to $\Delta_{\gamma \gamma}$ fulfills this condition on galactic scales, see Eq.~\eqref{CMBsmall}, we can recast Eq.~\eqref{neglect} as an upper limit on $\omega$:
\begin{equation}
\label{energyconstraint}
\omega \lesssim 700~\mbox{TeV}~\left(\frac{\mbox{10~kpc}}{L_\text{corr}} \right) \cdot \left(\frac{6~\mu \mbox{G}}{B_\text{T}} \right)^2 \; ,
\end{equation}
below which we can employ the simplified conversion probability of Eq.~\eqref{conversiongen}.

At higher energies, when Eq.~\eqref{energyconstraint} does not hold, we can estimate the conversion probability by considering $N_\text{corr}=L/L_\text{corr}$ domains in which the transverse magnetic field has the same value $B_{{\text T},{\text corr}}$. Then, the conversion probability is given by
\begin{equation}
P_{{\text g} \rightarrow \gamma} \simeq \frac{2N_\text{corr} B^2_{{\text T},\text{corr}} L^2_\text{corr}\sin^2 \left(\frac{1}{2}  \Delta_{\gamma \gamma} L_\text{corr} \right)  }{M^2_{{\text P}}\left| \Delta_{\gamma \gamma} L_\text{corr} \right|^2} \; .
\end{equation}
Averaging over \(\Delta_{\gamma\gamma} L_\text{corr}\gg1\) we find
\begin{equation}
\label{suppression}
P_{{\text g} \rightarrow \gamma} \simeq \frac{N_\text{corr} B^2_{{\text T},{\text corr}}}{M^2_\text{P} \Delta^2_{\gamma \gamma}} \; .
\end{equation}
Because the conversion probability decays as $P_{{\text g} \rightarrow \gamma} \propto 1/\omega^2$, no efficient graviton-to-photon conversion is expected in the case of ultra-high energies, $\omega \gg 1~\mbox{PeV}$.

% --  --  --  --  --  --  --  --  --  --  --  --  --  --  --  --  --  --  --  --  --  --  --  --  --  -- -
\section{The \(\gamma\)-ray flux}\label{sec:flux}
% --  --  --  --  --  --  --  --  --  --  --  --  --  --  --  --  --  --  --  --  --  --  --  --  --  -- -

The flux of gravitons per solid angle is related to the spectral energy density $d\rho_\text{gw}/d\ln \omega$ by 
\begin{equation}
\Phi_\text{gw}(\omega) = \frac{1}{4\pi} \cdot \frac{d\rho_\text{gw}}{d\ln \omega} \; .
\end{equation}
A small fraction of gravitons is converted into photons through the inverse Gertsenshtein effect:
\begin{equation}
\label{fluxgen}
\Phi_{\gamma} (\omega, \vec{n})=\Phi_\text{gw} (\omega) \cdot P_{{\text g} \rightarrow \gamma}(\omega, \vec{n})  \; ,
\end{equation}
where the conversion probability $P_{{\text g} \rightarrow \gamma}(\omega, \vec{n})$ depends on the direction $\vec{n}$ through the inhomogeneous magnetic field $\vec{B}(\vec{x})$. Because we are interested in \(\gamma\)-rays propagating within the Milky Way, we do not include the attenuation due to the interaction with the extragalactic background light (EBL).
%The factor $e^{-\tau (D, \omega)}$, %where $\tau (D, \omega)$ \fede{choose \%(L\) or \(D\)} is the optical depth, %takes into account the attenuation of %the photon flux due to the interaction %with the EBL through the process of %electron-positron pair creation \%(\gamma_{VHE}+\gamma_{EBL} \rightarrow %e^{+}+e^{-}\). Notice that the %attenuation is irrelevant in the Milky %Way, but should be taken into account %for distances $D \gg 10~\mbox{kpc}$. 
Restricting ourselves to graviton energies $\omega \lesssim 1~\mbox{PeV}$, cf. Eq.~\eqref{energyconstraint}, we can use Eq.~\eqref{simplified} for the conversion probability, which we substitute into Eq.~\eqref{fluxgen} to obtain
\begin{equation}
\Phi_{\gamma} (\omega, \vec{n}) \simeq \frac{3H^2_0 \cdot \Omega_\text{gw} }{8\pi} \left| \int^L_0 dz' \vec{B}_\text{T} \right|^2   \; , 
\end{equation}
where the graviton relic density is
\begin{equation}
\Omega_\text{gw} \equiv \frac{1}{\rho_\text{tot}} \cdot \frac{d\rho_\text{gw}}{d\ln \omega} \; ,
\end{equation}
and $\rho_\text{tot}=3H^2_0 M^2_\text{P}$ with $H_0$ the Hubble constant.
%The energy spectrum of converted photons follows that of the incoming gravitons $d\rho_\text{gw}/d\ln \omega$. Note that for $\omega \gtrsim 1~\mbox{PeV}$, the energy dependence should be modified to $d\rho_\text{gw}/(\omega^2 \cdot d\ln \omega)$ according to Eq.~\eqref{suppression}. This faster falloff with $\omega$ explains why we limit to the energies $\omega \lesssim 1~\mbox{PeV}$.
In this work we assume that the graviton spectrum, and hence also the resulting photon spectrum, is sharply peaked at some energy \(\omega\) (we provide a practical example of this in Sec.~\ref{sec:dm}), but our results are readily generalised to any spectral shape.

\begin{figure}
    \centering
    \includegraphics[width=0.8\textwidth]{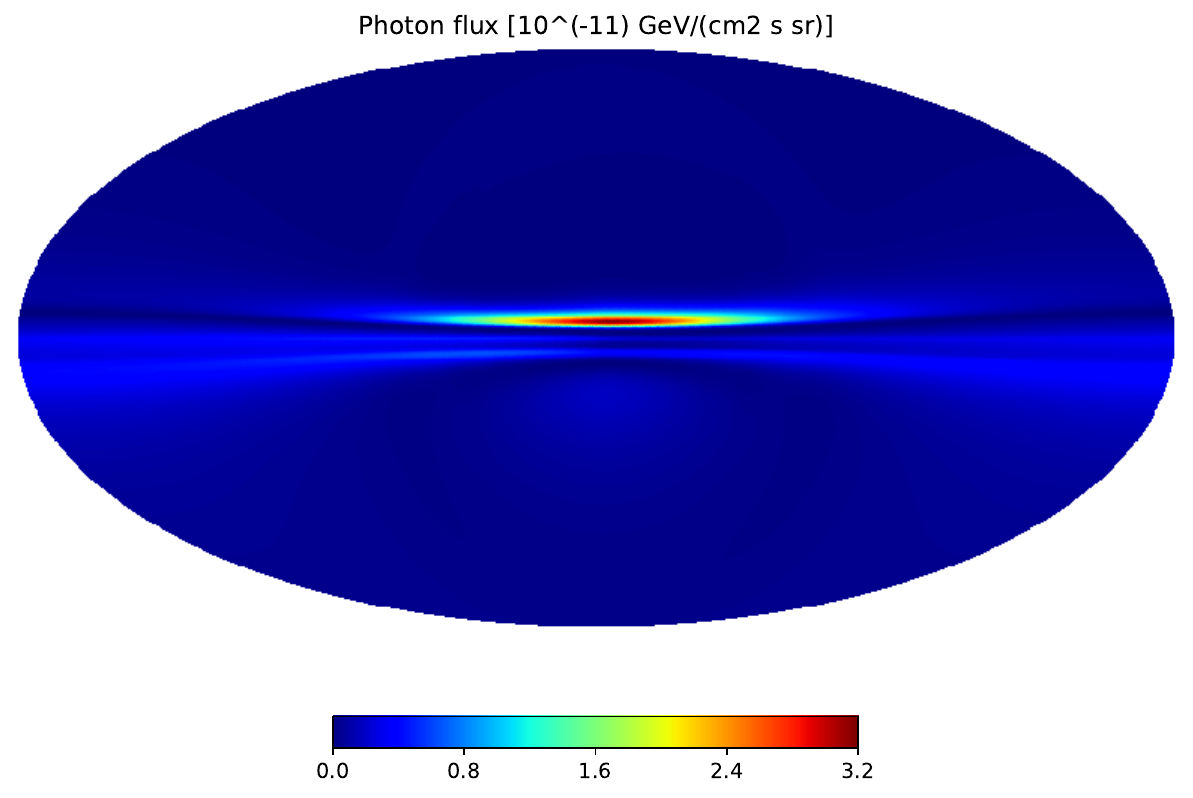}\\
    \includegraphics[width=0.8\textwidth]{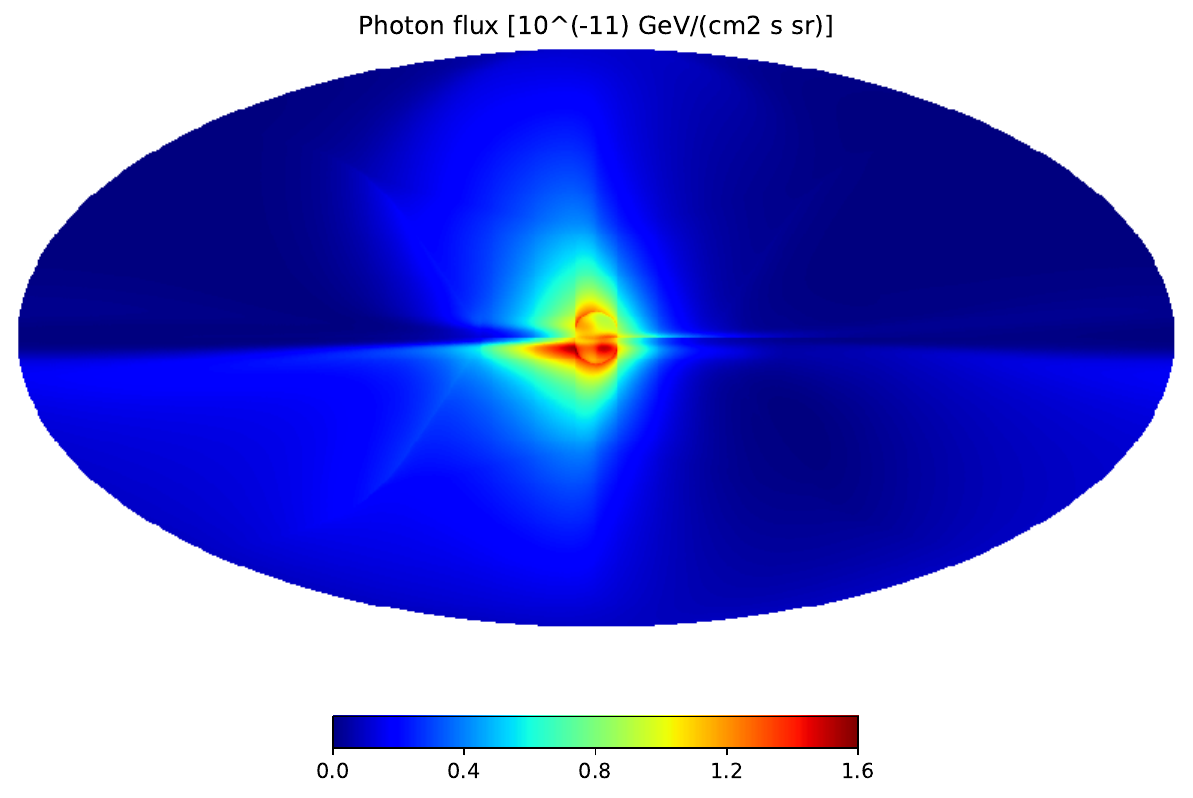}
    \caption{All-sky flux map for the Galactic magnetic field model of~\cite{Pshirkov:2011um} (top) and~\cite{GMF} (bottom), assuming \(\Omega_{gw}h_0^2=1\)---notice the different colour scale. The flux maps are in Mollweide projection with lines of sight starting 30 kpc from the Earth. Galactic longitude increases to the left in the plots, while Galactic latitude increases vertically. The centre of both plots corresponds to the line of sight in the direction of the Galactic centre. We consider the ordered magnetic field to be vanishing outside a sphere of radius 20 kpc away from the Galactic centre.}
    \label{sensitivity}
\end{figure}

The Galactic magnetic field that enters the expression of the \(\gamma\)-ray flux, \eqref{fluxgen}, is generally modelled as a sum of two components: a large-scale coherent field and a small-scale turbulent one~\cite{Boulanger:2018zrk}. Because the turbulent magnetic fields in the Milky Way have comparable strength but much smaller correlation length than the coherent component, they contribute significantly less to the photon flux~\cite{Carenza:2021alz}: for simplicity, we neglect them in what follows. In order to quantify the photon flux converted in the Milky Way, we adopt the large-scale magnetic field models of Refs.~\cite{GMF} (augmented with a central constant infill of \(5\,\mu\)G as in Ref.~\cite{Day:2015xea}) and~\cite{Pshirkov:2011um}. Normalising the photon flux to typical conversion probability values obtained from these magnetic field models, we find:
\begin{equation}
\label{flux}
\Phi_{\gamma} (\vec{n})=3 \cdot 10^{-12} \cdot \Omega_\text{gw}  h^2_0 \cdot \frac{P_{{\text g} \rightarrow \gamma} (\vec{n})}{10^{-16}} \cdot
\frac{\mbox{GeV}}{\mbox{cm}^2~\mbox{s}~\mbox{sr}} \; .
\end{equation}
Note that the conversion probability and hence the resulting photon flux is strongly direction-dependent: it is maximal in the directions where gravitons cross regions with large magnetic fields over large distances. In both models~\cite{GMF} and~\cite{Pshirkov:2011um}, the maximal probability $P_{{\text g} \rightarrow \gamma} \sim {\cal O}(10^{-15})$ is due to the halo component of the magnetic field protruding out of the Galactic plane. This translates into a flux peaking at a few~\(\times10^{-11}\,~\mbox{GeV}/\left(\mbox{cm}^2~\mbox{s}~\mbox{sr} \right)\) for $\Omega_\text{gw}h^2_0=1$ (cf.\ Fig.\ref{sensitivity}).

The magnetic field in the Galactic centre is much stronger, with a lower limit of $B \gtrsim 50~\mu \mbox{G}$ at $400~\mbox{pc}$, and a best fit of $B \sim 100~\mu \mbox{G}$~\cite{Crocker:2010xc}. The flux of photons from the Galactic centre then reads: 
\begin{equation}
\label{fluxgc}
\Phi_{\gamma}=3 N_\text{corr} \cdot 10^{-12} \cdot \Omega_\text{gw}  h^2_0 \cdot \left(\frac{B_\text{T}}{100~\mu \mbox{G}} \right)^2 \cdot \left( \frac{L_\text{corr}}{100~\mbox{pc}} \right)^2\cdot 
\frac{\mbox{GeV}}{\mbox{cm}^2~\mbox{s}~\mbox{sr}} \; ,
\end{equation}
where $N_\text{corr} \equiv L/L_\text{corr}$, and the size of the central region is estimated as $L \sim 1~\mbox{kpc}$. Because the larger value of the magnetic field in the Galaxy Centre is partially compensated by the smaller correlation length of the field, we do not expect a sizeable improvement compared to the diffuse flux from other regions of the Milky Way. Nonetheless, since the structure and strength of the magnetic field in the Galactic Centre is still the subject of significant uncertainties, the estimate~\eqref{fluxgc} may yield a considerably larger flux compared to that shown in Fig.~\ref{sensitivity}.

In order to determine whether the sub-PeV photon flux generated by the conversion of gravitons is detectable, we compare the diffuse photon flux Eq.~\eqref{flux} with the sensitivity to a diffuse \(\gamma\)-ray flux of LHAASO as estimated in Ref.~\cite{Neronov:2019ncc}. The sensitivity peaks in the energy range $100~\mbox{TeV} - 1~\mbox{PeV}$ where it is \({\cal O}(10^{-10})\,~\mbox{GeV}/\left(\mbox{cm}^2~\mbox{s}~\mbox{sr} \right)\), beyond which the sensitivity (and, incidentally, also the conversion probability) decreases. The LHAASO sensitivity is determined by the diffuse \(\gamma\)-ray background generated by multi-PeV cosmic rays in the Galaxy, which can be estimated by extrapolating Fermi-LAT measurements in the TeV range -- this background is still subject to significant uncertainties and is strongly latitude-dependent, see also Ref.~\cite{Luque:2022buq}. We assume here that other \(\gamma\)-ray backgrounds are negligible in the energy range of interest; this is justified since observations give only an upper limit on the high-latitude diffuse \(\gamma\)-ray background, which reads $\Phi_{\gamma} \lesssim 10^{-9}~\mbox{GeV}/\left(\mbox{cm}^2~\mbox{s}~\mbox{sr} \right)$ in the $100~\mbox{TeV} - 1~\mbox{PeV}$ range~\cite{Neronov:2021ezg}. Lastly, if the sources of the PeV neutrinos observed by IceCube are extragalactic, the corresponding photon flux on Earth will be strongly suppressed because of the opacity of the EBL.

From the discussion above and Fig.~\ref{sensitivity}, it follows that the gravitationally induced $\gamma$-ray flux is below the peak sensitivity of LHAASO at \(100~\text{TeV}\) for $\Omega_\text{gw} h^2_0 < 1$. In particular, a realistic relic graviton abundance \(\Omega_\text{gw} h^2_0\sim 0.01\) as predicted by the scenario outlined in Sec.~\ref{sec:dm} is currently two orders of magnitude below the sensitivity reach. Nevertheless, future facilities with better sensitivity, together with a better modelling (or actual detection, as expected in the case of the cosmic-ray \(\gamma\)-ray background with LHAASO) of this background, we can expect that the shimmering graviton flux can be detected, in particular leveraging its directional dependence as well as peculiar energy spectrum that separates it from other diffuse backgrounds.

% --  --  --  --  --  --  --  --  --  --  --  --  --  --  --  --  --  --  --  --  --  --  --  --  --  -- -
\section{Gravitons from decaying dark matter}\label{sec:dm}
% --  --  --  --  --  --  --  --  --  --  --  --  --  --  --  --  --  --  --  --  --  --  --  --  --  -- -

Gravitons in the TeV - PeV range with a cosmologically relevant abundance can arise from the
%: i) the evaporation of ultra-light primordial black holes by Hawking radiation~\cite{Dolgov:2000ht, Anantua:2008am, Dolgov:2011cq}; ii) 
decay of heavy particles which have only gravitational interactions~\cite{Ema:2021fdz}.
%---Bremmstrahlung graviton emission falls in the latter category~\cite{Nakayama:2018ptw, Huang:2019lgd, Ghoshal:2022kqp}.
If this mechanism operates in the early Universe, i.e., before recombination, there is an inevitable bound on their abundance, $\Omega_\text{gw} h^2_0 \lesssim 10^{-6}$, as inferred from the Planck+BAO constraint on the effective number of neutrino species~\cite{Planck:2018vyg}.
%Furthermore, the initial fraction of ultra-light primordial black relative to radiation is the subject of rather stringent, albeit model-%dependent, constraints~\cite{Inomata:2020lmk, Papanikolaou:2020qtd, Domenech:2020ssp, Keith:2020jww}. \rs{ In 50, 51, the constraint on %initial fraction also comes from $\Delta N_{\rm eff}$ produced by gravitational waves (e.g., density fluctuation), but in low-frequency region.} 
On the other hand, the decay into gravitons may take place after recombination, in which case the bound $\Omega_\text{gw} h^2_0 \lesssim 10^{-6}$ does not apply.
%\footnote{In scenario I, evaporation of ultra-light primordial black holes may take place after recombination in a narrow range of masses %close to $M \sim 10^{15}~\mbox{g}$. In this case, however, we expect stringent limits from\(\gamma\)-rays emitted along with gravitons by black %holes as a part of Hawking radiation.} 
Heavy particles maintaining only gravitational interactions with other matter fields are natural candidates for dark matter. While most of the dark matter must be stable on timescales exceeding the age of the Universe, a small component is allowed to decay in a shorter time.

Let us consider a real singlet scalar field $\Psi$, which constitutes a fraction $f$ of the total dark matter density and which decays into gravitons with rate $\Gamma_{\Psi}$. We discuss two cases. 

{\it Case A:} $\Gamma_{\Psi} \gtrsim H_0$. In this case, the field \(\Psi\) has fully decayed by the present epoch, i.e., at some redshift $z_\text{dec}>0$, which implies that $f \ll 1$. The resulting graviton relic abundance is estimated as 
\begin{equation}
\nonumber 
\Omega_\text{gw} h^2_0 \simeq \frac{f \Omega_\text{dm} h^2_0}{1+z_\text{dec}} \; ,
\end{equation}
where $\Omega_\text{dm}h^2_0$ is the dark matter relic abundance. The peak frequency of produced gravitons reads
\begin{equation}
\nonumber 
\omega \simeq \frac{M_{\Psi}}{2(1+z_\text{dec})} \; .
\end{equation}
One reason to consider decaying dark matter is the possibility that it may alleviate the Hubble tension, see~\cite{Berezhiani:2015yta}. This requires $f \simeq 0.1$, cf.\ Refs.~\cite{Chudaykin:2017ptd, Bucko:2022kss}, corresponding to 
\begin{equation}
\Omega_\text{gw} h^2_0 \lesssim \frac{10^{-2}}{1+z_\text{dec}} \; ,
\end{equation}
where we substituted $\Omega_\text{dm}h^2_0=0.12$~\cite{Planck:2018vyg}. It is crucial that the heavy particles $\Psi$ primarily decay into gravitons, while the decay into the Standard Model is suppressed. Otherwise, the fraction $f$ is constrained to be $f \lesssim 10^{-5}-10^{-7}$ (depending on the decay channel)~\cite{Kalashev:2019xkw} and it would give a negligible contribution to a relic abundance of gravitons (nor would it help with the Hubble tension).

{\it Case B:} $\Gamma_{\Psi} \lesssim H_0$. In this case, the field $\Psi$ is stable on cosmological timescales and thus can constitute the entirety of the dark matter, i.e., one can set $f=1$. The limit on the lifetime of the field $\Psi$, i.e, $\tau_{\Psi} \simeq \Gamma^{-1}_{\Psi}$, reads $\tau_{\Psi} \gtrsim 10 \tau_\text{U}$~\cite{Audren:2014bca}, where $\tau_\text{U}$ is the age of the Universe\footnote{Notice that the lower bound here is presently subject to uncertainties: the constraint on the lifetime $\tau_{\Psi}$ from CMB alone is stronger by a factor two~\cite{Bucko:2022kss}, 
while the $\sigma_8$ tension is alleviated with $\tau_{\Psi} \sim 5 \tau_\text{U}$~\cite{Enqvist:2015ara}.}. Consequently, the relic abundance of gravitons in this picture is bounded as
\begin{equation}
\label{limitstable} 
\Omega_\text{gw} h^2_0 \lesssim 0.01 \; ,
\end{equation}
while the peak frequency is given by $\omega \approx M_{\Psi}/2$. 
%Notice that this mechanism suggests that the energy density of the gravitons produced by the decay is strongly peaked around 
%\(\omega_{peak}\), in which case the constraint~\eqref{limitstable} can be slightly relaxed \fede{how much?}. We nevertheless keep the more %conservative limit \eqref{limitstable} in what follows.

The decay of the scalar $\Psi$ into a pair of gravitons can take place through its coupling to quadratic curvature invariants:
\begin{equation}
\label{int}
\frac{\Psi}{\Lambda} \cdot R^2 \,, \qquad \frac{\Psi}{\Lambda} \cdot R_{\mu \nu} R^{\mu \nu } \,, \qquad \frac{\Psi}{\Lambda} \cdot R_{\mu \nu \lambda \rho} R^{\mu \nu \lambda \rho}  \; ,
\end{equation}
where $R$, $R_{\mu \nu}$,~$R_{\mu \nu \lambda \rho}$ are the Ricci scalar, Ricci tensor, and Riemann tensor, respectively; $\Lambda$ is the energy scale at which these operators become relevant. Notice that the coupling $\Psi R$ does not lead to the decay into gravitons: this becomes evident upon switching to the Einstein frame, where such a coupling is absent. The interactions~\eqref{int} lead to the decay rate~\cite{Ema:2021fdz}:
\begin{equation}
\label{ratepair}
\Gamma_{\Psi} \simeq \frac{M^7_{\Psi}}{4\pi \Lambda^2 M^4_{\text{P}}} \; ,
\end{equation}
where $M_{\Psi}$ is the mass of the field $\Psi$. For the values $\Lambda \simeq M_{\text{P}}$ and $\Lambda \simeq M_{\Psi}$, the condition $\Gamma_{\Psi} \simeq 0.1~\tau^{-1}_\text{U}$, which yields the largest possible abundance of gravitons in this case, gives 
$M_{\Psi} \simeq 10^{10}~\mbox{GeV}$ and $M_{\Psi} \simeq 2~\mbox{PeV}$, respectively. Hence, sub-PeV gravitons are generated for values of $\Lambda$ slightly below $M_{\Psi}$.

One important comment is in order here. While the coupling $\sim \Psi R$ does not lead to the decay into a pair of gravitons, it nevertheless triggers the production of (beyond) the Standard Model particles with a rate that is also inversely proportional to the fourth power of the Planck mass~\cite{Watanabe:2006ku} and thus can exceed the rate~\eqref{ratepair} in some range of parameters. If this is the case, the constraints on the \(\gamma\)-ray cascade initiated by the decay forces the fraction of superheavy dark matter to be $f \lesssim 10^{-5}\mbox{--}10^{-7}$~\cite{Kalashev:2019xkw} (in case A) or the dark matter lifetime to exceed the age of the Universe by $\sim 10$ orders of magnitude~\cite{Kachelriess:2018rty} (in case B). In both cases, gravitons are produced with a negligible abundance. Our discussion in this section assumes that the coupling $\sim \Psi R$ leads to a negligible decay rate, which can always be adjusted by a proper choice of the coupling constant.

% --  --  --  --  --  --  --  --  --  --  --  --  --  --  --  --  --  --  --  --  --  --  --  --  --  -- -
\section{Discussion}\label{sec:disc}
% --  --  --  --  --  --  --  --  --  --  --  --  --  --  --  --  --  --  --  --  --  --  --  --  --  -- -

In this work we have suggested that the highest energy for which a cosmological abundance of gravitons can be detected through their conversion into photons in the external electromagnetic field, is in the sub-PeV range. This is the case because for $\omega \gg 1~\mbox{PeV}$ the graviton-to-photon conversion is blocked by QED effects, i.e., the effective photon mass acquired in the magnetic field. We have shown that the largest possible $\gamma$-ray flux caused by gravitons is about two orders of magnitude below the sensitivity of LHAASO to a diffuse $\gamma$-ray flux for energies of $\omega \sim 100~\mbox{TeV}\text{--} 1~\mbox{PeV}$. Future improvements, owing to a better understanding of backgrounds, a larger effective areas and longer observation times, can open the opportunity to detect cosmological sub-PeV shimmering gravitons.

For our estimates of the $\gamma$-ray flux in Sec.~\ref{sec:flux}  we have focussed exclusively on the magnetic field of the Milky Way. This is justified because the propagation length of sub-PeV gravitons is limited to be in the $\mbox{kpc}\text{--}\mbox{Mpc}$ range due to EBL absorption. Moreover, our results are robust against a possible input from extragalactic magnetic fields~\cite{Neronov:2010gir}, because their comoving strength at 1~Mpc is constrained to be $B \lesssim 1~\mbox{nG}$, whether they are primordial~\cite{Jedamzik:2018itu} or not ~\cite{Pshirkov:2015tua}, so that they give a negligible contribution to the flux. Starburst galaxies host magnetic fields as large as $B \sim 100~\mu \mbox{G}$~\cite{Thompson:2006is}, but there are only a few of them within a Mpc radius, and their flux is low because of their small conversion length compared to their distance. % ratio are seen under very small angles.
For the same reasons, we neglect the contribution from regions with strong intracluster magnetic fields.

As shown in Fig.~\ref{sensitivity}, the predicted $\gamma$-ray flux is strongly sensitive to the choice of the Milky Way magnetic field model. While current state-of-the-art models lead to an overall agreement on the maximal flux (within a factor of two), they predict drastically different direction-dependences. Placing this direction-dependence on firm grounds as well as better understanding the magnetic field in the Galactic bulge is crucial to discriminate between gravitationally-induced and cosmic-ray-triggered $\gamma$-ray backgrounds.\\

{\bf Acknowledgments.} We are indebted to Mikhail Kuznetsov and Alexander Vikman for useful discussions. S.~R., R.~S., and F.~U. acknowledge the support from the European Structural and Investment Funds and the Czech Ministry of Education, Youth and Sports (Project CoGraDS -CZ.02.1.01/0.0/0.0/15003/0000437). The work of G.~T. is supported by the Grant Agency of the Czech Republic, GA\v CR grant 20-28525S. This article/publication is based upon work from COST Action COSMIC WISPers CA21106, supported by COST (European Cooperation in Science and Technology). We wish to express our gratitude to Error Bar and Pivnice Satanka, where part of this work was carried out.

\appendix

\section{Details of graviton-to-photon conversion}\label{app:details_conv}

The goal of this Appendix is to obtain the matrix~\eqref{ggamma} that describes the graviton-to-photon conversion. Before we set off, notice that the effective photon mass does not play a role in this calculation, so we can set it to zero. In other words, we can ignore the Euler-Heisenberg correction to the electromagnetic action.

The equations of motion describing the evolution of gravitons and photons in the external magnetic field $\vec{B}^{e}$ are given by
\begin{align}
    &\Box h_{ij}=-\frac{\sqrt{2}}{M_{P}}\left(B_{i}B^{\text{e}}_{j}+B^{\text{e}}_{i}B_{j}\right) \; , \\
    &\Box A_{i}=\frac{\sqrt{2}}{M_\text{P}}\partial_{j}h_{ik}F^{\text{e}}_{kj} \; .
\end{align}
We assume that the graviton field obeys the gauge conditions $h_{0i}=0$, $\partial_{i}h_{ij}=0$ and $h^{i}_{i}=0$, while the gauge field fulfills $A_{0}=0$ and $\partial_{i}A_{i}=0$. We also ignored spatial derivatives of the external magnetic field, which is plausible  given its large correlation length. Using the identity $F_{kj}=-\epsilon_{kjl}B_{l}$ and expanding the graviton and photon fields into the sum over polarisations using Eqs. \eqref{Aexp} and \eqref{hexp}, we rewrite the system above as
\begin{align}
\label{system}
&\sum_{\lambda =\times, +} (\omega^{2}+\partial_{k}^{2})h_{\lambda} e^{\lambda}_{ij}=-\frac{\sqrt{2}}{M_\text{P}}\sum_{\lambda=\parallel, \bot} \left[\epsilon_{ilk}(\partial_{l}A_{\lambda}) {\epsilon}^{\lambda}_{k}B^{\text{e}}_{j}+\epsilon_{jlk}(\partial_{l}A_{\lambda}) {\epsilon}^{\lambda}_{k}B^{\text{e}}_{i} \right] \; , \nonumber \\
&\sum_{\lambda=\parallel, \bot}(\omega^{2}+\partial_{k}^{2})A_{\lambda} \epsilon^{\lambda}_{i}=-\frac{\sqrt{2}}{M_\text{P}}\sum_{\lambda =\times, +}  (\partial_{j}h_{\lambda})e^{\lambda}_{ik}F^{\text{e}}_{kj} \; .
\end{align}
We assume the gravitational/electromagnetic waves to be propagating along the $z$-direction, i.e.,
\begin{equation}
    h_{+,\times}(z)=\tilde{h}_{+,\times}e^{ikz} \; , \qquad  A_{\parallel,\bot}(z)=\tilde{A}_{\parallel,\bot}e^{ikz} \; .
\end{equation}
In the high-energy regime we are interested in, we can write 
\begin{equation}\label{Boxapprox}
(\omega^{2}+\partial_{k}^{2})(\cdot)=(\omega-i\partial_{k})(\omega+i\partial_{k})(\cdot)=(\omega+k)(\omega+i\partial_{k})(\cdot) \approx 2\omega (\omega+i\partial_{k})(\cdot)  \; .
\end{equation}
%\begin{align}\label{basiscr%ossproducts}
%\left((\Vec{\nabla}A_{\para%llel})\times\vec{\epsilon}_%%{\parallel}\right)\cdot\vec{\epsilon}_{\bot}&=ik\Tilde{A}_{\parallel}\left(\hat{e}_{z}\times\hat{e}_{x}\right)\cdot\hat{e}_{y}=ik\Tilde{A}_{\parallel} \nonumber \\
%\left((\Vec{\nabla}A_{\bot})\times\vec{\epsilon}_{\bot}\right)\cdot\vec{\epsilon}_{\parallel}&=ik\Tilde{A}_{\bot}\left(\hat{e}_{z}\times\hat{e}_{y}\right)\cdot\hat{e}_{x}=-ik\Tilde{A}_{\bot}
%\end{align}
 Using Eq. \eqref{Boxapprox} and the relations $e^{\lambda}_{ij} e^{\lambda'}_{ij}=2\delta_{\lambda \lambda'}$, $\epsilon^{\lambda}_i \epsilon^{\lambda'}_i=\delta_{\lambda \lambda'}$, we recast the system~\eqref{system} into a system of four first-order differential equations describing the evolution of gravitons and photons with given polarisations:
\begin{align}
 &(i\partial_{z}+\omega)h_{+} = \frac{i}{\sqrt{2} M_\text{P}}\biggl[\left(\vec{B}^{\text{e}}\cdot\vec{\epsilon}_{\bot}\right)A_{\parallel}+
 \left(\vec{B}^{\text{e}}\cdot\vec{\epsilon}_{\parallel}\right)A_{\bot}\biggr] \; , \\
 &(i\partial_{z}+\omega)h_{\times} =\frac{i}{\sqrt{2}M_\text{P}}\biggl[\left(\vec{B}^{\text{e}}\cdot\vec{\epsilon}_{\bot}\right)A_{\bot}-\left(\vec{B}^{\text{e}}\cdot\vec{\epsilon}_{\parallel}\right)A_{\parallel}\biggr] \; ,\\ 
 &(i\partial_{z}+\omega)A_{\parallel}  =\frac{i}{\sqrt{2} M_\text{P}}\left[\left(\vec{B}^{\text{e}}\cdot\vec{\epsilon}_{\parallel}\right)h_{\times}-\left(\vec{B}^{\text{e}}\cdot\vec{\epsilon}_{\bot}\right)h_{+}\right] \; , \\ 
 &(i\partial_{z}+\omega)A_{\bot} =-\frac{i}{\sqrt{2} M_\text{P}}\left[\left(\vec{B}^{\text{e}}\cdot\vec{\epsilon}_{\parallel}\right)h_{+}+\left(\vec{B}^{\text{e}}\cdot\vec{\epsilon}_{\bot}\right)h_{\times}\right] \; .
\end{align}
This reproduces Eq.~\eqref{ggamma}, where we omitted the superscript $\text{e}$.
The results of this Appendix agree with the ones of Ref.~\cite{Ejlli:2018hke}.

\end{document}